\documentclass[10pt,twocolumn,superscriptaddress,preprintnumbers, showpacs]{revtex4}%
\usepackage{epsf}
\usepackage{amsmath}
\usepackage{amssymb}
\usepackage{epsfig}
\usepackage{latexsym}
\usepackage{amsfonts}
\usepackage{varioref}
\usepackage{ifthen}
\usepackage{graphicx}
\usepackage{amssymb,amsmath,amsthm}%
\providecommand{\U}[1]{\protect\rule{.1in}{.1in}}
\begin{document}
\parindent 0mm
\parindent 0mm
\setlength{\parskip}{\baselineskip}
\setcounter{page}{1}
\mbox{ }
\title{Determination of the temperature dependence of the up- down-quark mass in QCD}
\author{C. A. Dominguez} 
\author{L. A. Hernandez}
\affiliation{Centre for Theoretical \& Mathematical Physics, and Department of Physics, University of Cape Town,
Rondebosch 7700, South Africa}
\date{\today}
\begin{abstract}
\noindent 
The temperature dependence of the sum of the QCD  up- and down-quark masses, $(m_u + m_d)$ and the pion decay constant, $f_\pi$, are determined from two thermal finite energy QCD sum rules for the pseudoscalar-current correlator. This quark-mass  remains mostly constant for temperatures well below the critical temperature for deconfinement/chiral-symmetry restoration. As this critical temperature is approached, the quark-mass increases sharply with increasing temperature. This increase is far more pronounced if the temperature dependence of the pion mass (determined independently from other methods) is taken into account. The behavior of $f_\pi(T)$ is consistent with the expectation from chiral symmetry, i.e. that it should follow the thermal dependence of the quark condensate, independently of the quark mass. 
\end{abstract}
\maketitle
\parindent 0mm \setlength{\parskip}{\baselineskip} \thispagestyle{empty}
\pagenumbering{arabic} 
\noindent 
The method of QCD sum rules (QCDSR) \cite{QCDSR1} is a well established technique to obtain results in QCD analytically. In particular, it has been widely used to determine the values of all quark masses \cite{reviewqmass}-\cite{FLAG}, except for the top-quark. This is achieved e.g. in the light-quark sector by considering QCD sum rules for the pseudoscalar current correlator, proportional to the square of the quark masses. Current precision matches that from e.g. lattice QCD (LQCD)\cite{FLAG}. The extension of QCDSR to finite temperature was first proposed in \cite{BS}, and subsequently used over the years in a plethora of applications. Of particular relevance are the thermal QCDSR results obtained in the light-quark  axial-vector \cite{axial}, and  vector channel \cite{vector}, which will be used here.
The most appropriate correlation function in the determination of quark masses is the pseudoscalar current correlator
\begin{equation}
\psi_{5}(q^{2})= i\int d^{4}x \,  e^{i q x} \, <0| T(\partial^\mu A_\mu(x) \, \partial^\nu A^\dagger_\nu(0))|0> ,
\end{equation}
with
\begin{equation}
\partial^\mu A_\mu(x) = m_{ud} : \overline{d}(x)\, i \,\gamma_5\, u(x): \;,
\end{equation}
and the definition
\begin{equation}
m_{ud} \equiv (m_u+md) \, \simeq \, 10 \,{\mbox{MeV}} \,,
\end{equation}
where $m_{u,d}$ are the  quark masses in the $\overline{MS}$-scheme at a scale $\mu = 2  \, {\mbox{GeV}}$ \cite{reviewqmass}-\cite{FLAG}, and $u(x)$, $d(x)$, the corresponding quark fields. The numerical value of these quark masses at $T=0$ is irrelevant, as we shall only determine ratios. This has been the standard procedure in thermal QCDSR, always at one-loop order, ever since their introduction \cite{BS}.
The relation between the QCD and the hadronic representation of current correlators is obtained by invoking Cauchy's theorem in the complex square-energy plane, Fig.1, which leads to the finite energy sum rules (FESR) \cite{QCDSR1}-\cite{reviewqmass}
\begin{equation}
\int_{\mathrm{0}}^{s_0} ds\, \frac{1}{\pi}\, Im \,\psi_5(s)|_{HAD}  = -\frac{1}{2 \pi i}  \oint_{C(|s_0|) }\, ds  \,\psi_5(s)|_{QCD} \, , 
\end{equation}
\begin{eqnarray}
 \int_{\mathrm{0}}^{s_0} \frac{ds}{s}\, \frac{1}{\pi}\, Im \,\psi_5(s)|_{HAD}  &+& \frac{1}{2 \pi i}  \oint_{C(|s_0|) }\, \frac{ds}{s}  \,\psi_5(s)|_{QCD} \nonumber  \\ [.3cm]
& =& \psi_5(0) \;,
\end{eqnarray}
where 
\begin{equation}
\psi_5(0)= {\mbox{Residue}}\, [\psi_5(s)/s]_{s=0} \,.
\end{equation}
The radius of the contour, $s_0$, in Fig.1 is large enough for QCD to be valid on the circle. Information on the hadronic spectral function on the left hand side of Eq.(4) allows to determine the quark masses entering the contour integral. Current precision determinations of quark masses require the introduction of integration kernels on both sides of Eq.(4). These kernels are used to enhance or quench hadronic contributions, depending on the integration region, and on the quality of the hadronic information available. They also deal with the issue of potential quark-hadron duality violations, as QCD is not valid on the positive real axis in the resonance region. This will be of no concern here, as we are going to determine only  ratios, e.g. $m_{ud}(T)/m_{ud}(0)$, to leading order in the hadronic and the QCD sectors. This has been so far the standard approach in thermal FESR.\\  
 \begin{figure}[ht]
 \begin{center}
   \includegraphics[height=.22\textheight]{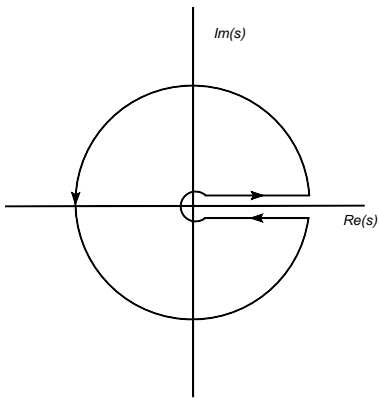}
   \caption{Integration contour in the complex s-plane. The discontinuity across the real axis brings in the hadronic spectral function, while integration around the circle involves the QCD correlator. The radius of the circle is $s_0$, the onset of QCD.}     
    \label{fig:figure1}
 \end{center}
 \end{figure}
To this order, the QCD expression of the pseudoscalar correlator, Eq.(1), is
\begin{eqnarray}
&&\psi_5(q^2)|_{QCD} =  m_{ud}^2 
\left\{- \frac{3}{8 \,\pi^2} q^2  \ln\left(\frac{- q^2}{\mu^2}\right)\right.  \nonumber  \\ [.3cm]
 &+& \left.  \frac{m_{ud} \;\langle \bar{q}q \rangle}{q^2} - \frac{1}{8\,  q^2} \;\langle \frac{\alpha_s}{\pi} G^2 \rangle + {\cal{O}} \left( \frac{O_6}{q^4} \right)\right\} ,
\end{eqnarray} 
where $\langle \bar{q} q \rangle = (- 267 \pm 5 \, {\mbox{MeV}})^3$ from \cite{GMOR},  and $\langle \frac{\alpha_s}{\pi} G^2  \rangle = 0.017 \pm 0.012 \, {\mbox{GeV}}^4$ from \cite{GG}.
The  gluon- and quark-condensate contributions in Eq.(7) are, respectively, one and two orders of magnitude smaller than the leading perturbative QCD term. Furthermore, at finite temperature both condensates decrease with increasing $T$, so that they can be safely ignored in the sequel.\\
The QCD spectral function at finite $T$, obtained from the Dolan-Jackiw formalism \cite{DJ}, in the rest frame of the medium $(q^2 = \omega^2 - |\bf{q}|^2 \rightarrow \omega^2)$ is given by
\begin{equation}
{\mbox{Im}}\; \psi_5(q^2,T)|_{QCD}= \frac{3}{8 \, \pi} m_{ud}^2 (T) \; \omega^2  \left[ 1 \,-\, 2 \, n_F(\omega/ 2T) \right],
\end{equation}
where $n_F(x) = (1 + e^{x})^{-1}$ is the Fermi thermal factor.  At finite temperature there is in principle an additional contribution \cite{BS} from a cut centred at the origin in the complex energy $\omega$-plane with extension $- |{\bf q}| \leq \omega \leq + |{\bf q}|$. In the rest frame of the thermal  medium ($|{\bf{q}}| \rightarrow 0$), this cut can either lead to a vanishing contribution to a spectral function, or to a delta function of the energy, $\delta(\omega)$, depending on the correlator. If present, this so-called QCD scattering term is proportional to the squared temperature times the delta function $\delta(\omega)$. In the case of the pseudoscalar correlator, Eq.(1), this term is absent. This is due to the overall factor of $q^2$ in the perturbative QCD term in Eq.(7), which prevents the formation of a delta function $\delta(\omega)$. A non-vanishing QCD scattering term enters in e.g. the correlator of vector and axial-vector currents, thus differentiating them from the pseudoscalar correlator. Such a term  also appears in the hadronic representation of a current correlator, and it involves hadron loops. For instance, in the case of the vector current correlator, the hadronic scattering term is due to a two-pion loop. In the hadronic sector of the pseudoscalar correlator the scattering term is due to a phase-space suppressed two-loop three-pion contribution, which is negligible in comparison with the pion-pole term
\begin{equation}
{\mbox{Im}}\; \psi_5(q^2,T)_{HAD}=
2 \, \pi\,f_\pi^2(T) \, M_\pi^4(T) \; \delta(q^2-M_\pi^2)\,,
\end{equation}
where $f_\pi = 92.21 \, \pm \, 0.02 \,{\mbox{MeV}}$ \cite{PDG2014}.
\begin{figure}[h!]
 \begin{center}
   \includegraphics[height=.22\textheight]{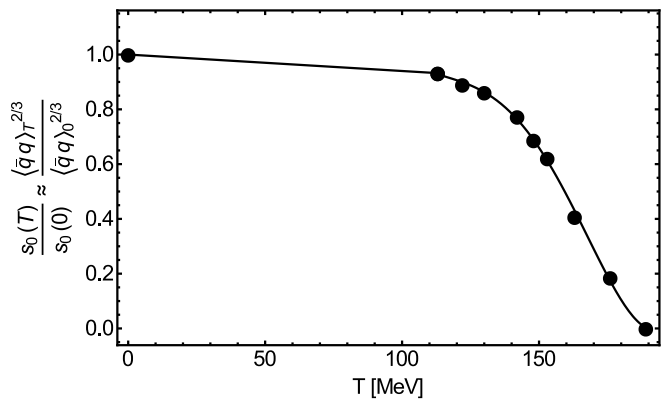}
   \caption{The thermal quark condensate normalized to its value at $T=0$ from \cite{LQCD} (solid circles), and  our fit (solid line). The phenomenological deconfinement parameter, $s_0(T)/s_0(0)$, is expected to follow the quark condensate behaviour \cite{axial}, except possibly very close to $T_c$.}
 \label{fig:figure2}
 \end{center}
 \end{figure}
 Corrections to this relation arise from the radial excitations of the pion, e.g. $\pi(1300)$, and $\pi(1800)$. In the chiral $SU(2) \times SU(2)$ symmetry limit, these states are not Goldstone bosons, so that their decay constants vanish in this limit. In the real world these decay constants are at the level of a few $MeV$. Their contribution to the pseudoscalar correlator is only meaningful in precision determinations of the light quark masses from QCD FESR \cite{reviewqmass}. Furthermore, we find that the value of $s_0$ at $T=0$ from the FESR is below these resonances, whose already large width $(\Gamma \simeq 200 - 600 \, {\mbox{MeV}})$ will grow even larger with increasing temperature. Together with the well  established fact that $s_0(T)$ decreases monotonically with increasing $T$, these states can  be safely neglected here. We also notice that at the end we shall divide all thermal results by their $T=0$ values.\\
\begin{figure}[ht]
 \begin{center}
   \includegraphics[height=.24\textheight]{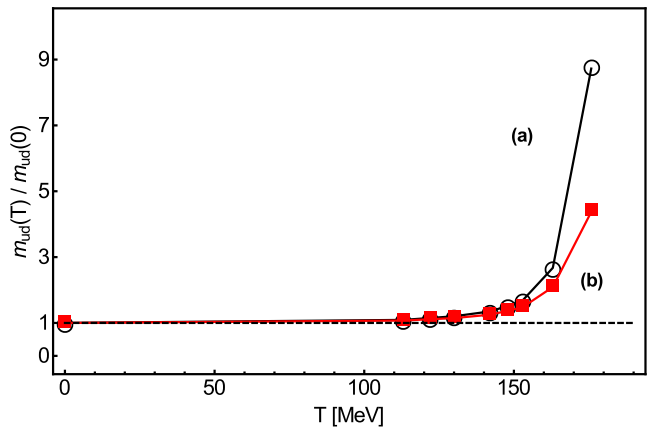}
   \caption{The ratio of the quark masses $m_{ud}(T)/m_{ud}(0)$ as a function of $T$ from the FESR Eqs.(10)-(11). Curve (a) is for a $T$-dependent pion mass from \cite{MPIT}, and curve (b) is for a constant pion mass.}
 \label{fig:figure2}
 \end{center}
 \end{figure}
 \begin{figure}[ht]
  \begin{center}
    \includegraphics[height=.24\textheight]{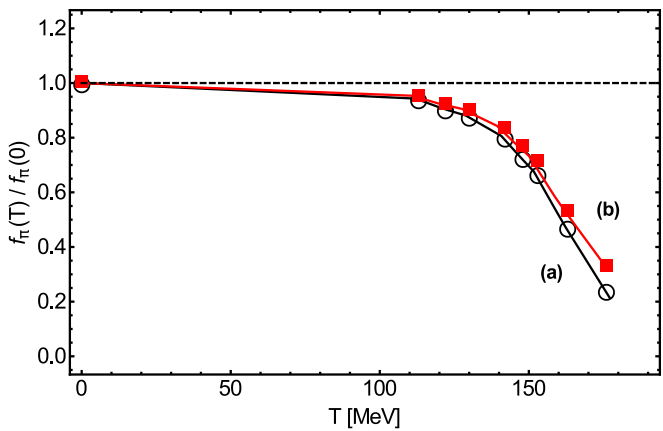}
    \caption{The ratio of the pion decay constant $f_\pi(T)/f_\pi(0)$ as a function of $T$ from the FESR Eqs.(10)-(11). Curve (a) is for a $T$-dependent pion mass from \cite{MPIT}, and curve (b) is for a constant pion mass.}
  \label{fig:figure2}
  \end{center}
  \end{figure}
  
The two FESR, Eqs.(3)-(4), at finite $T$ become
\begin{eqnarray}
2 f_\pi^2(T) M_\pi^4(T) = \frac{3 \,m_{ud}^2(T)}{8 \pi^2}\int_0^{s_0(T)} s \left[1-2 n_F \left(\frac{\sqrt{s}}{2 T}\right)\right]ds , \nonumber\\
\end{eqnarray}
\begin{eqnarray}
2 \,f_\pi^2(T) \, M_\pi^2(T) &=& - 2 \,m_{ud}(T) \,\langle\bar{q} q\rangle(T) \, + \, \frac{3}{8 \pi^2} \,m_{ud}^2(T) \nonumber  \\ [.3cm]
&\times& \int_0^{s_0(T)} \left[1 - 2 n_F \left(\frac{\sqrt{s}}{2 T}\right)\right] \,ds. 
\end{eqnarray}
Equation (11) is the thermal Gell-Mann-Oakes-Renner relation incorporating a higher order QCD quark-mass correction, ${\cal{O}} (m^2_{ud})$. While at $T=0$ this correction is normally neglected \cite{GMOR}, at finite temperature this cannot be done, as it is of the same order in the quark mass as the right-hand-side of Eq.(10). As done in deriving Eq.(10), hadronic corrections due to radial excitations of the pion have been neglected in  Eq.(11).\\
The thermal quark condensate is the order parameter of chiral-symmetry restoration, i.e. the phase transition between a Nambu-Goldstone   to a Wigner-Weyl realization of $SU(2) \times SU(2)$. On the other hand, $s_0(T)$ is a phenomenological parameter signalling quark deconfinement \cite{BS}. One expects these two parameters to be related if the two phase transitions take place at a similar critical temperature. In fact, the relation
\begin{equation}
\frac{s_0(T)}{s_0(0)}\simeq \left[\frac{\langle \bar{q} q \rangle(T)}{\langle \bar{q} q \rangle(0)}\right]^{2/3},
\end{equation}
was suggested long ago from FESR in the axial-vector channel \cite{CAD0}, and confirmed soon after by a more detailed analysis \cite{Gatto}. Notice that while Eq.(12) only involves ratios, one still matches the dimensions of the individual parameters.  Using current information this relation was 
reconfirmed in \cite{axial}, using thermal FESR for the vector-current correlator (independent of the pseudoscalar correlator, Eq.(1)).  Further support for the relation, Eq.(12), is provided by LQCD results \cite{LQCD2}. We do not expect this relation to be valid very close to the critical temperature, $T_c$, as we are using the thermal quark condensate for finite quark masses, which is non-vanishing close to $T_c$.  Using this result on $s_0(T)/s_0(0)$ as input in the FESR, Eqs. (10)-(11), together with  LQCD results for $\langle \bar{q} q\rangle(T)$ for finite quark masses \cite{LQCD}, and independent determinations of $M_\pi(T)$ \cite{MPIT}, we can determine the ratios $m_{ud}(T)/m_{ud}(0)$ and $f_\pi(T)/f_\pi(0)$. 
We expect the latter ratio to be close to $\langle \bar{q} q\rangle(T)/\langle \bar{q} q\rangle(0)$. This is because in the Nambu-Goldstone realization of chiral symmetry the pion mass vanishes as the quark mass
\begin{equation}
 M_\pi^2 = B \, m_q \;,
\end{equation}
while the pion decay constant vanishes as the quark condensate
\begin{equation}
f_\pi^2 = \frac{1}{B} \langle\bar{q} q \rangle
\end{equation}
with $B$ a constant \cite{CHPT}. This expectation is confirmed by the results from the FESR as discussed next.\\
The LQCD results for the thermal quark condensate \cite{LQCD} are shown in Fig.2 (solid circles), together with our fit to these points (continuous curve). Results from the FESR, Eqs.(10)-(11), are shown in Figs. 3 and 4, respectively. The thermal quark mass is essentially constant at low temperatures, rising sharply at $T \simeq 150\,{\mbox{MeV}}$. This rise is far more pronounced for the case of a $T$-dependent pion mass (obtained from \cite{MPIT}). Beyond $T \simeq 170 \;{\mbox{MeV}}$ the FESR cease to have real solutions, as $s_0(T)$ approaches zero. Figure 4 confirms the expectation from chiral symmetry that $f_\pi(T)/f_\pi(0)$ should be independent of  the thermal behaviour of the pion mass, and follow instead the behavior of the quark condensate, Eq.(14).\\
The temperature behaviour of the quark mass determined here is consistent with the expectation that at the critical temperature for deconfinement the free quarks would acquire a constituent mass, much bigger than the small QCD mass.
\begin{center}
{\bf Acknowledgements}
\end{center}
This work was supported in part by the National Research Foundation (South Africa).


\begin{thebibliography}{99}                                      \bibitem{QCDSR1} For a review see e.g. P. Colangelo and A. Khodjamirian, in: "At the Frontier of Particle Physics/ Handbook of QCD", M. Shifman, ed. (World Scientific, Singapore 2001), Vol. 3, 1495-1576.        
\bibitem{reviewqmass} For a recent review see e.g. C. A. Dominguez,  Int. J. Mod. Phys. A {\bf 29}, 1430069 (2014).
\bibitem{FLAG} S. Aoki {\it{et al}}, FLAG Coll. arXiv:13108555.
\bibitem{BS} A. I. Bochkarev and M. E. Shaposnikov, Nucl. Phys. B {\bf{268}}. 220 (1986).
\bibitem{axial} C. A. Dominguez, M. Loewe, and Y. Zhang, Phys. Rev. D {\bf 86}, 034030 (2012).
\bibitem{vector} A. Ayala, C. A. Dominguez, M. Loewe, and Y. Zhang, Phys. Rev. D {\bf 86}, 114036 (2012).
\bibitem{GMOR} J. Bordes, C. A. Dominguez, P. Moodley, J. Pe\~{n}arrocha, and K. Schilcher, J. High Energy Phys. {\bf 05}, 064 (2010).
\bibitem{GG} C. A. Dominguez, L. A. Hernandez, K. Schilcher, and H. Spiesberger, J. High Ener. Phys. {\bf 03}, 053 (2015). 
\bibitem{DJ} L. Dolan and R. Jackiw, Phys. Rev D {\bf 9 }, 3320 (1974).
\bibitem{PDG2014} K. A. Olive {\it et al.} (Particle Data Group), Chinese Phys. {\bf 38}, 090001 (2014).
\bibitem{CAD0} C. A. Dominguez and M. Loewe, Phys. Lett. B {\bf 233}, 201 (1989).
\bibitem{Gatto} A. Barducci, R. Casalbuoni, S. de Curtis, R. Gatto, and G. Pettini, Phys. Lett. B {\bf 244}, 311 (1990).
\bibitem{LQCD2}S. Borsanyi, {\it et al.}, J. High Energy Phys. {\bf 09}, 073 (2010); T. Bhuttacharya, {\it et al.}, HotQCD Coll., Phys. Rev. Lett. {\bf 113}, 082001 (2014).
\bibitem{LQCD} G. S. Bali, F. Bruckmann, G. Endr\"{o}di, Z. Fodor, D. Katz, and A. Sch\"{a}fer, Phys. Rev. D {\bf 86}, 071502 (2012).
\bibitem{MPIT} M. Heller and M. Mitter, arXiv:1512.05241.
\bibitem{CHPT}J. Gasser , and H. Leutwyler, Nucl. Phys. B {\bf 250}, 465 (1985); Phys. Lett. B {\bf 184}, 83 (1987).
\end{thebibliography}
\end{document}